\documentclass[twocolumn,3p]{elsarticle}

\usepackage{lineno,hyperref,amsmath,amssymb,graphicx}
\usepackage{graphics}
\journal{Journal of \LaTeX\ Templates}

\begin{document}

\begin{frontmatter}

\title{Spin-1 Bose-Hubbard model with two- and three-body interactions}

\author[unal]{A. F. Hincapie-F}
\author[unal]{R. Franco}
\author[unal]{J. Silva-Valencia\corref{cor1}}
\ead{jsilvav@unal.edu.co}
\address[unal]{Departamento de F\'{\i}sica, Universidad Nacional de Colombia, A.A. 5997, Bogot\'a-Colombia.}

\begin{abstract}
We investigated the ground state of spin-1 bosons interacting under local two- and three-body interactions in one dimension by means of the 
density matrix renormalization group method. We found that the even-odd asymmetry will be obtained or not depending on the relative values of 
the two-  and  three-body interactions. The Mott insulator lobes are spin isotropic, the first showing a dimerized pattern and the second being 
composed of singlets. The three-body interactions disfavor a longitudinal polar superfluid and a quantum phase transition to a transverse 
polar superfluid occurs, which could be continuous or discontinuous.
\end{abstract}

\begin{keyword}
\texttt{Bose-Hubbard model}\sep degenerate gases \sep quantum phase transitions
\MSC[2010] 00-01\sep  99-00
\end{keyword}

\end{frontmatter}


\section{Introduction\label{intro}}
The macroscopic occupation of a unique quantum state was observed by confining alkalai atoms with magneto-optical traps, opening a new branch of the 
physics called ultracold atoms, which has linked and promoted several areas of physics~\cite{IBloch-08}. The possibility of exploring diverse physical 
phenomena led to the development of new cooling and trapping techniques, in particular new ways to confine atoms without freezing the spin degree of freedom, 
which was achieved by creating purely optical traps with lasers~\cite{Stamper-RMP13}. Using the above technique to confine sodium or rubidium atoms, 
it has been possible to observe a Bose-Einstein condensate for each hyperfine state~\cite{StamperKurn-PRL98}, Larmor precession~\cite{Higbie-PRL05}, 
spin domains~\cite{Stenger-Nature98}, coherent spin dynamics~\cite{Chang-NP05}, and spontaneous symmetry breaking ~\cite{Sadler-Narure06}, among other 
phenomena. Note that cold-atom setups constitute a new means of studying quantum magnetism~\cite{Greif-Science13} and the interplay between charge and 
spin degrees of freedom~\cite{Gorshkov-NP10}.\par 
Cold-atom setups are highly tunable, allowing the experimental observation of a reversible quantum phase transition between a superfluid state and a 
Mott insulator one for bosonic atoms in magneto-optical traps by Greiner {\it et al.}~\cite{Greiner-N02a}. To observe the transition, one needs to 
efficiently control the ratio of interactions to the mobility of atoms by increasing the depth of an optical lattice, manipulating interactions with Feshbach 
resonance~\cite{Chin-RMP10}, or periodically shaking the lattice~\cite{Lignier-PRL07}. The superfluid-Mott insulator transition for spinless bosons is 
second-order and has been described in terms of the Bose-Hubbard model, which considers spinless bosons that can jump between the sites of a lattice and that 
interact through a local repulsion term~\cite{Kuhner-PRB98,Ejima-EL11}. Spinor boson systems also undergo a superfluid-Mott insulator transition, but 
this can be first (second) order for antiferromagnetic~\cite{Stamper-RMP13} (ferromagnetic ~\cite{Kimura-PRL05}) spin interactions, and a quadratic Zeeman 
energy, in addition to the above methods, can be used to carry out the transition~\cite{Jiang-PRA16}.\par 
A minimal model for describing S=1 bosons confined in an optical lattice must consider a kinetic term and a local two-body repulsion term in way similar to 
the spinless Bose-Hubbard model, but from experiments we know that the magnetic properties are crucial. Therefore, an extra term it is necessary, and a 
local spin-dependent interaction was suggested, leading to the S-1 Bose-Hubbard Model~\cite{Imambekov-PRA03}. The local spin-dependent interaction can be 
tuned using optical Feshbach resonance~\cite{Theis-PRL04,Thalhammer-PRA05}, and the nature of this interaction can be antiferromagnetic ($^{23}$Na) or 
ferromagnetic ($^{87}$Rb), depending on the relative magnitudes of the scattering lengths in the quintuplet and singlet channels~\cite{Stamper-RMP13}. When 
the on-site spin-spin interaction is antiferromagnetic, the S-1 Bose-Hubbard Model exhibits an even-odd asymmetry in the Mott lobes; i.e. for an even global 
density in the system the Mott lobes grow as the spin exchange interaction parameter increases, while odd global density decreases. The magnetic 
properties of the Mott insulating lobes are very interesting. The odd lobes exhibit a dimerized order, while the even lobes exhibit competition between a nematic 
phase and a spin singlet one~\cite{Yip-PRL03,Imambekov-PRL04,Tsuchiya-PRA04,Rizzi-PRL05,Kimura-PRL05,Bergkvist-PRA06,Apaja-PRA06,Pai-PRB08,Toga-JPSJ12,
Kimura-PRA13,Natu-PRB15,Li-PRA16}. For a ferromagnetic on-site spin-spin interaction, the asymmetry discussed above is absent. The Mott 
lobes always decrease as the spin exchange interaction parameter increases, regardless of the global density. In addition, the magnetic properties are 
described by a ``long-range'' ferromagnetic order~\cite{Batrouni-PRL09}.\par 
Confining $^{87}$Rb atoms in optical lattices, it was possible to observe evidence of the effects of multi-body interactions using an atom interferometric 
technique~\cite{Will-N10} or photon-assisted tunneling experiments~\cite{Ma-PRL11}. This experimental evidence of multi-body interactions stimulated several 
theoretical studies on the effect of taking into account multi-body interactions in spinless boson 
chains~\cite{JSV-PRA11,JSV-EPJB12,Sowinski-PRA12,Ejima-PRA13,Dutta-RPP15} and the emergence of new proposals where the three-body interactions are 
dominant~\cite{Daley-PRA14,Paul-PRA16}. To understand the experimental results, taking into account the spin degree of freedom, Mahmud and Tiesinga made a 
calculation using perturbation theory and derived effective spin-dependent and spin-independent three-body interaction potentials~\cite{Mahmud-PRA13}.
Recently, the ground state of a spinor gas under {\it only} three-body interactions was studied numerically, finding that the even-odd asymmetry of the 
two-body antiferromagnetic chain is absent and the density drives first-order superfluid-Mott insulator transitions for even and odd 
lobes~\cite{Hincapie-PRA16}.\par 
The phase diagram of spin-1 bosons under two- and three-body interactions was calculated by Nabi and Basu using mean-field 
theory~\cite{Nabi-ArX16}. They found that  as the area of the Mott lobes increases with respect to the superfluid region, the critical point of the 
superfluid-Mott insulator transition moves to larger values of the hopping with the strength of the three-body interaction term, and they concluded that 
for antiferromagnetic interactions, the even-odd asymmetry remains. Motivated by these results, we decided to take a step beyond the 
mean-field  and study the S-1 Bose-Hubbard Hamiltonian with two- and three-body interactions using the density matrix renormalization group method. 
We found that for the parameters of Nabi and Basu the even-odd asymmetry does not occur, and this is retained for larger values of the two-body 
spin-dependent interaction. We also observed a quantum phase transition from a longitudinal polar superfluid to a transverse polar superfluid due to the 
three-body interactions.\par 
The rest of the article is organized as follows: In Sec. 2 we set up the spin-1 Bose-Hubbard model with two- and three-body interactions, and the atomic 
limit is considered. We present our main numerical results in Sec. 3 based on the density matrix renormalization group method. We summarize our results 
in Sec. 4.\par

\section{Model\label{model}}
Taking into account the interaction terms between spin-1 bosons proposed by Imambekov {\it et al.}~\cite{Imambekov-PRA03} and 
Mahmud {\it et al.}~\cite{Mahmud-PRA13}, a degenerate gas of spin-1 atoms can be described by the following effective Hamiltonian
\begin{eqnarray}\label{Hgen} 
\mathcal{H}=-t \sum_{i,\sigma} \left( b^{\dagger}_{i,\sigma} b_{i+1,\sigma}+  
\text{h.c} \right) +\nonumber \\ \sum_i\left\{ \hat{n}_i(\hat{n}_i-1)\left[\frac{U_0}{2}+\frac{V_0}{6}(\hat{n}_i-2)\right]\right\}+ \nonumber \\ 
\sum_i\left\{(\mathbf{\hat{S}}_i^2-2\hat{n}_i)\left[ \frac{U_2}{2}+ \frac{V_2}{6}(\hat{n}_i-2)\right]\right\}, 
\end{eqnarray}
\noindent where  $b^{\dagger}_{i,\sigma}(b_{i,\sigma})$ creates (annihilates) a boson with spin component $\sigma$ on site $i$ of a one-dimensional 
optical lattice of size $L$, the total number of particles at the site $i$ is $\hat{n}_i=\sum_{\sigma}n_{i,\sigma} $ and  
$n_{i,\sigma} = b^{\dagger}_{i,\sigma}b_{i,\sigma}$. Representing the spin-1 Pauli matrices by $\mathbf{T}_{\sigma,\sigma '}$, the local spin 
operator is $\mathbf{\hat{S}}_i=\sum_{\sigma,\sigma '}b^{\dagger}_{i,\sigma}\mathbf{T}_{\sigma,\sigma '} b_{i,\sigma '}$.\par  
The physical meaning of each term in the Hamiltonian (\ref{Hgen}) is: The first term describes the kinetic energy of bosons with $t$ being the 
tunneling force between nearest sites. The interactions between bosons appear in the remaining terms; specifically, the spin-independent terms and 
the spin-dependent interactions are shown in the second and third line, respectively. The parameter $U_s$ represents the two-body interaction in 
the $S$ channel given by $U_0= 4\pi \hbar^2(a_0+2a_2)/3M$ and $U_2= 4\pi\hbar^2(a_2-a_0)/3M$ where $a_s$ are the scattering lengths for $S=0$ 
and $S=2$ channels and $M$ is the mass of the atom~\cite{Stamper-RMP13}. In the present paper, we consider $U_2>0$, i.e. antiferromagnetic 
interaction. The three-body interaction parameters are $V_0$ and $V_2$. The validity of the perturbation theory imposes $V_0<<U_0$ and 
$V_2=2\frac{U_2}{U_0}V_0$~\cite{Mahmud-PRA13}.\par  
At the atomic limit of the Hamiltonian (\ref{Hgen}) ($t \rightarrow 0$), the energy is given by 
\begin{eqnarray}
E_M(n_i)=n_i(n_i-1)\left[\frac{U_0}{2}+\frac{V_0}{6}(n_i-2)\right]+ \nonumber \\
\left(<\mathbf{\hat{S}}_i^2>-2n_i\right)\left[\frac{U_2}{2}+\frac{V_2}{6}(n_i-2)\right],
\end{eqnarray}
\noindent $<\mathbf{\hat{S}}_i^2>=S(S+1)$ being the expected value of the local spin operator, which has a very important role. It is zero if the 
number of particles at each site is even, due to the singlets formation. If the number of particles at each site is odd $<\mathbf{\hat{S}}_i^2>=2$, and 
one boson remains free while the others form a singlet. Therefore we have to distinguish between the cases of an even and an odd number of particles, and the 
energy when the number of particles is even ($n_e$) is given by
\begin{eqnarray}
E_M(n_e)=n_e(n_e-1)\left[\frac{U_0}{2}+\frac{V_0}{6}(n_e-2)\right]- \nonumber \\n_e\left[U_2+\frac{V_2}{3}(n_e-2)\right] ,
\end{eqnarray}
\noindent while for an odd number of particles ($n_o$)
\begin{eqnarray}
E_M(n_o)=n_o(n_o-1)\left[\frac{U_0}{2}+\frac{V_0}{6}(n_o-2)\right]+ \nonumber \\ \left(1-n_o\right)\left[U_2+\frac{V_2}{3}(n_o-2)\right].
\end{eqnarray}
The absence of the kinetic term leads to a ground state with an integer number of interacting particles per site; hence the ground state is a  
Mott insulator one. The boundaries of the Mott lobes at the atomic limit are given by $\mu(n\rightarrow n+1)= E(n+1)-E(n)$, which corresponds to the 
chemical potential under the canonical ensemble. Taking into account the above definition, we obtain 
$\mu(n_o \rightarrow n_o+1)=U_0n_o-2U_2+\frac{V_0}{2}n_o(n_o-1)+V_2(1-n_o) $ and 
$\mu(n_e \rightarrow n_e+1)= U_0n_e+\frac{V_0}{2}n_e(n_e-1)-\frac{V_2}{3}n_e $. From the above expressions, we found that for a finite nonzero 
$V_0$ value, the odd lobes decrease as the two-body spin interaction parameter ($U_2$) grows, and note that these lobes do not vanish when $U_2$ 
reaches its maximum value $U_2/U_0=0.5$. Also, the upper boundary of the Mott lobes with even filling is independent of $U_2$. Now, if we fix  
$U_2$, we obtain that the boundaries of the first Mott lobe are independent of the three-body interactions, a fact that we expected, because we 
do not have fluctuations. As we increase the density, the three-body spin-independent interaction parameter has an important role in increasing 
the area of the Mott lobes.\par 
\begin{figure}[t] 
\includegraphics[width=18pc]{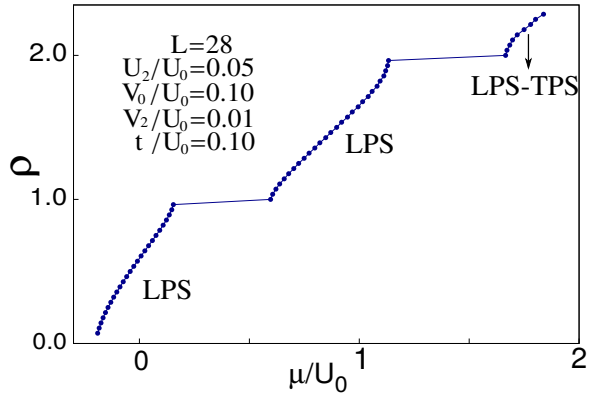} 
\caption{\label{fig1} Global density $\rho$  as a function of the chemical potential $\mu$ for a finite chain of $L=28$ sites. The plateau 
indicates the possibility that for integer densities Mott insulator phases occur. Two different superfluid regions appear: longitudinal polar 
superfluid (LPS) and transverse polar superfluid (TPS). The parameters here are:$t/U_0=0.10$, $U_2/U_0=0.05$, $V_0/U_0 =0.10$, and $V_2/U_0 =0.01$. 
The lines are visual guides.} 
\end{figure} 
\section{Numerical results\label{Res}}
As the hopping parameter is turned on ($t\neq 0$), the particles can be delocalized, and quantum fluctuations emerge in the system. Due to this, the 
ground state will change, and we expect a ground state with delocalized particles, when the kinetic energy will be the dominant energy scale. Therefore, 
the Hamiltonian (\ref{Hgen}) exhibits different ground states, and we need to distinguish between the diverse phases of the model. A first option 
is to calculate the charge gap at the thermodynamic limit. For finite lattices of size $L$, the charge gap is given by 
$\Delta(L)=\mu^p(L)-\mu^h(L) =E_0(L,S^z,N+1)+E_0(L,S^z,N-1) -2 E_0(L,S^z,N)$ where $\mu^p(L)(\mu^h(L))$ is the chemical potential to add (remove) a 
particle, and $E_0(L,S^z,N)$ is the ground state energy of a lattice with $N$ spin-1 bosons and $S^z$ projection of the total spin. We use the density 
matrix renormalization group (DMRG) method with open boundary conditions to calculate the ground state for lattices of a size up to 
$L=128$~\cite{White-PRL92}.  The dimension of the local Hilbert space basis is fixed by choosing a maximum occupation number $\hat{n}_{max}=5$ to 
guarantee accurate results. In the DMRG procedure, we kept up to $m=350$ states per block and obtained a discarded weight around $10^{-5}$ in the
worst case, and the ground-state energy converges to an absolute error of $10^{-3}$ or better. We set our energy scale choosing $U_0=1.0$ and 
studied the  ground-state for various values of the interacting parameters considering that  $U_2<0.5U_0$, $V_0<<U_0$ and 
$V_2=2\frac{U_2}{U_0}V_0$.\par
We want to emphasize that our DMRG code reproduces the already-known results of the limit cases with only two- or three-body interactions for 
spinless and spinful bosons. For instance, when $U_2=V_2=0$, the Hamiltonian (\ref{Hgen}) describes spinless bosons under two- and three-body 
interactions. We observed that our phase diagram (not shown) matches those reported by several authors for one-dimensional spinless bosons with 
only two- or three-body interactions~\cite{JSV-EPJB12,Sowinski-CEJP14}. Of course the known spinful results reported by 
Rizzi {\it et al.}~\cite{Rizzi-PRL05}, Batrouni {\it et al.}~\cite{Batrouni-PRL09}, and Hincapie {\it et al.}~\cite{Hincapie-PRA16} were 
verified.\par 
\begin{figure}[t] 
\includegraphics[width=18pc]{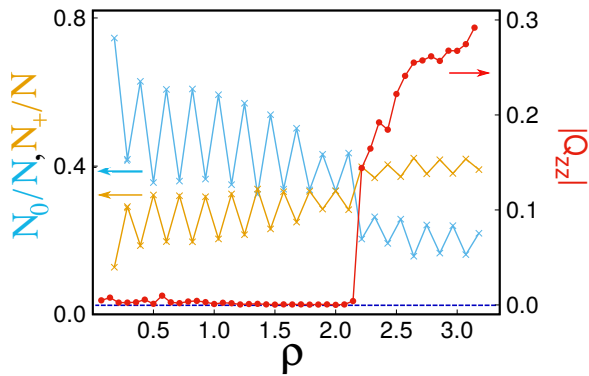} 
\caption{\label{fig2} The spin population fraction in each spin projection ($N_+/N$ and  $N_0/N$) and the $z$-component of the nematic order parameter
$Q_{zz}$ as a function of the global density $\rho$ for a finite chain of $L=28$ sites, a fixed hopping $t/U_0=0.10$ and interaction parameters 
$U_2/U_0=0.05$, $V_0/U_0 =0.10$, and $V_2/U_0 =0.01$. The lines are visual guides.} 
\end{figure} 
We took into account the theorems proposed by Katsura {\it et al.}\cite{Katsura-PRL13}, and all the calculations presented in this paper were done in 
the sector with $S^z=0$.\par
The evolution of the ground state as the number of particles increases is shown in Fig.~\ref{fig1} for a finite chain with $L=28$ sites, a fixed 
hopping $t/U_0=0.10$, and interaction parameters $U_2/U_0=0.05$, $V_0/U_0 =0.10$, and $V_2/U_0 =0.01$. The above parameters were chosen taking into 
account a recent mean-field calculation about spinor bosons with two- and three-body interactions~\cite{Nabi-ArX16}. We observe that the chemical 
potential increases monotonously as the global density grows, but at integer densities a plateau appears, which indicates that a finite energy gap 
was found for integer densities and the ground state corresponds to an incompressible Mott insulator at the thermodynamic limit. Far away from the integer densities, we do not 
obtain an energy gap. Therefore, the ground state is compressible and the spinor bosons are delocalized throughout the lattice. We found that the 
ground state can be either superfluid or Mott insulator. Note that the length of the plateau for the global density $\rho= N/L=2$ is larger than the one 
for $\rho=1$, which is due to two-body spin-interaction. From Fig.~\ref{fig1}, it is clear that the compressibility 
$\kappa=\partial \rho /\partial \mu$ is always positive, 
a fact that suggests the absence of first-order phase transitions from superfluid to Mott insulator~\cite{Batrouni-PRL00}. In 2009, 
Batrouni {\it et al.} reported a first-order transition at the second lobe of a spinor systems with only two-body interactions in spite of 
$\kappa=\partial \rho /\partial \mu > 0$~\cite{Batrouni-PRL09, Forges-PRB13}. For a spinor chain under only three-body interactions, we showed that 
the phase transitions are of the first-order kind for even and odd Mott lobes~\cite{Hincapie-PRA16}. In order to explore what happens in a one-dimensional 
system of spin-$1$ bosons under two- and three-body interactions, we calculated the spin population fraction in each spin projection 
($N_+/N$ and  $N_0/N$. The component $N_-/N$ is not necessary due to $N_+/N=N_-/N$) as a function of the global density, and the results are shown 
in Fig.~\ref{fig2}. Taking into account that the wave function of a singlet state is 
$|2,0\rangle=\sqrt{2/3}|1,-1\rangle|1,1\rangle-\sqrt{1/3}|1,0\rangle|1,0\rangle$; in this state, we must obtain that $N_+/N=N_-/N=N_0/N$, which 
happens in the range $1.5\lesssim \rho \lesssim 2.1$. However, for densities different from $\rho=2$, singlets bound states of two spinor 
bosons try to form themselves for a finite lattice, this being exactlly achieved at the thermodynamic limit only with  $\rho=2$. These results are similar to those 
reported by Batrouni {\it et al.} However, as the density increases, we observe a quantum phase transition from a longitudinal polar superfluid 
($N_0>N_+$) to a transverse polar superfluid ($N_0<N_+$), which happens due to the three-body  interactions between the spin-1 bosons. Comparing these 
results with our previous report about spinor bosons with only three-body interactions~\cite{Hincapie-PRA16}, we observe that the critical density for 
which the quantum transition from longitudinal polar to transverse polar superfluid occurs will depend on the two- and three-body interactions.\par
\begin{figure}[t] 
\includegraphics[width=18pc]{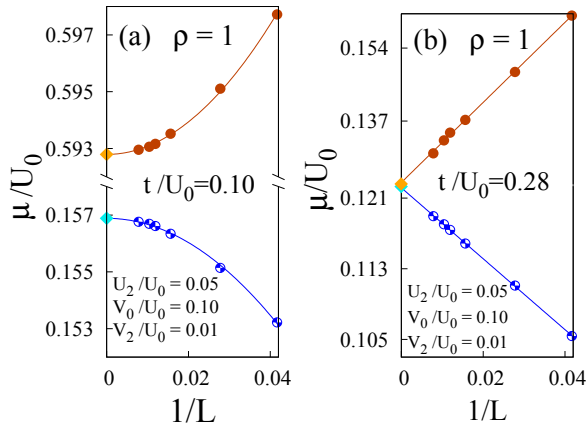} 
\caption{\label{fig3}System size dependence on the chemical potential. In both panels, the upper set of data corresponds to the energy to add a particle, 
and the lower one to the energy to remove one. As the hopping parameter increases, the ground state passes from a gapped state ($t/U_0=0.10$) to a 
gapless one ($t/U_0=0.28 $). The extrapolated values of the chemical potential at the thermodynamic limit are represented by diamonds. 
The lines are visual guides.} 
\end{figure} 
\begin{figure}[t!]
\begin{minipage}{18pc}
\includegraphics[width=18pc]{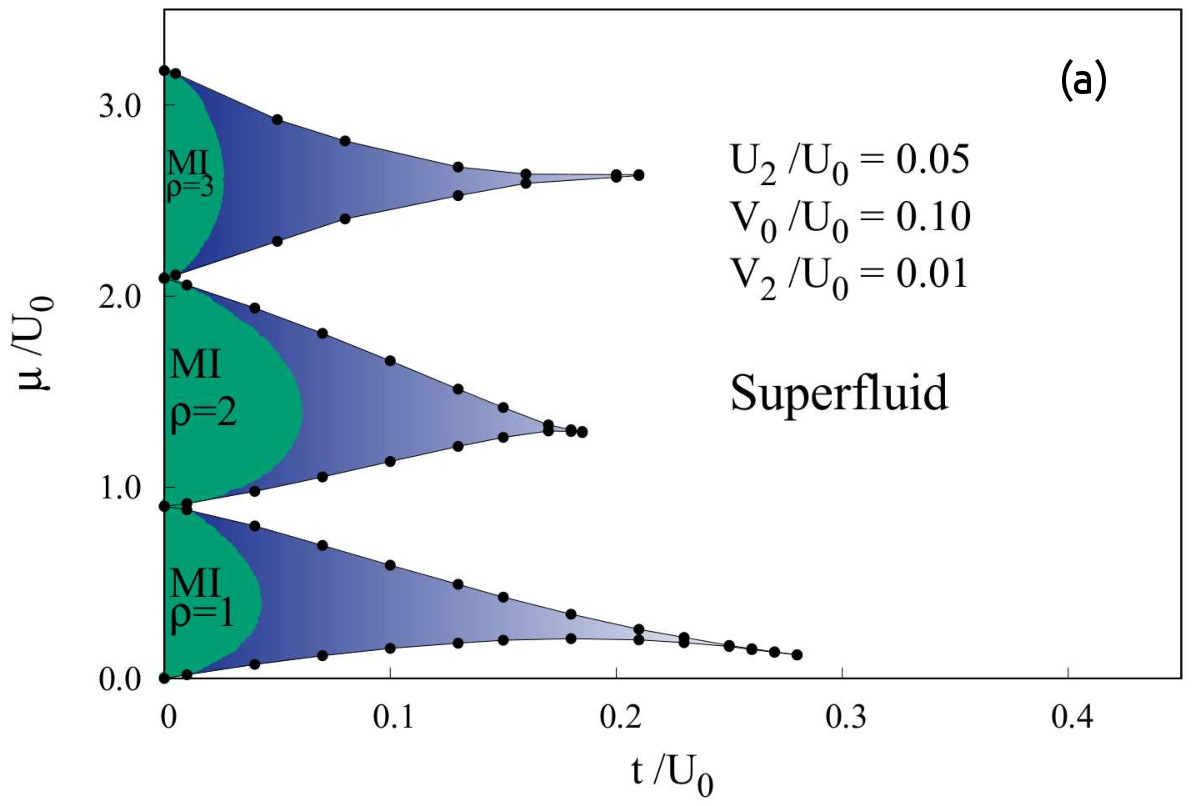}
\end{minipage}
\hspace{5pc}%
\begin{minipage}{18pc}
\includegraphics[width=18pc]{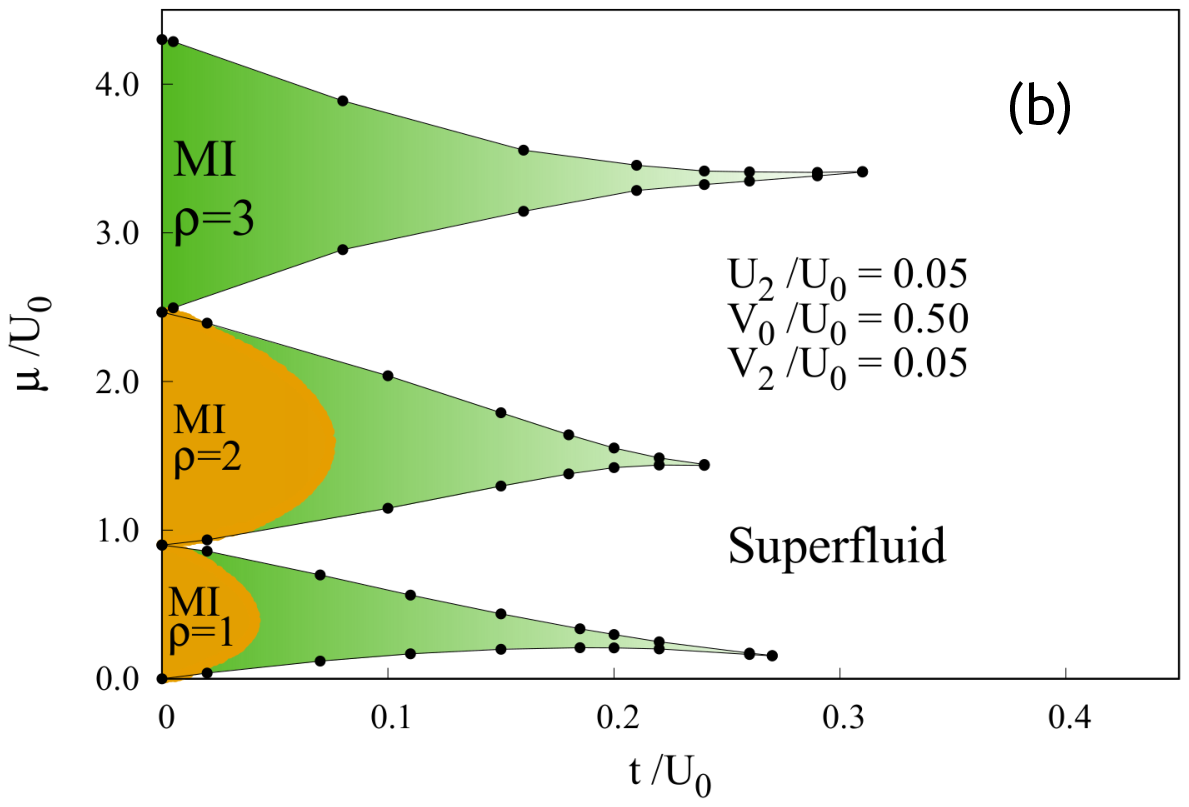}
\end{minipage}
\hspace{5pc}%
\begin{minipage}{18pc}
\includegraphics[width=18pc]{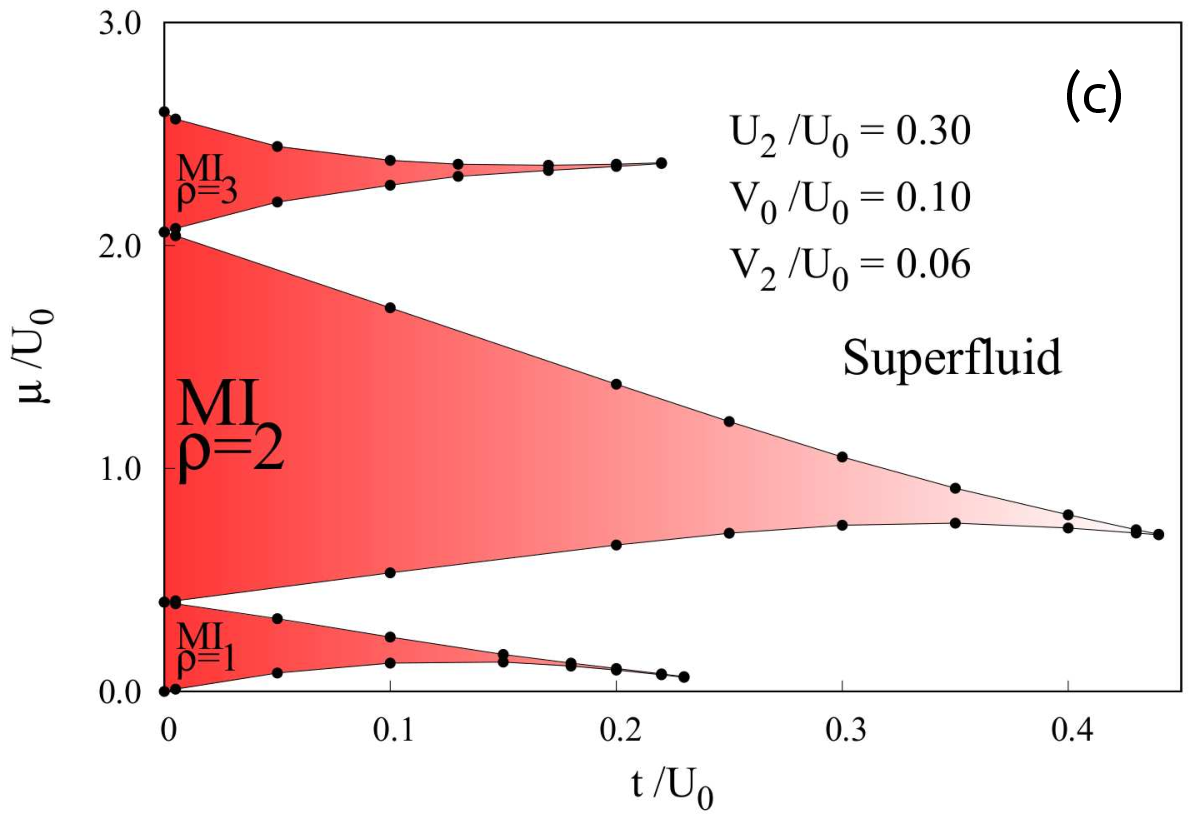}
\caption{\label{fig4}Phase diagram for the first three Mott lobes of the spin-1 Bose-Hubbard model with two- and three-body interactions and 
antiferromagnetic spin interaction (a) $U_2/U_0=0.05$, $V_0/U_0=0.1$, and $V_2/U_0=0.01$ (b)  $U_2/U_0=0.05$, $V_0/U_0=0.5$  and  $V_2/U_0=0.05$. 
(c) $U_2/U_0=0.3$, $V_0/U_0=0.1$, and $V_2/U_0=0.06$. The points are extrapolation values at the thermodynamic limit from DMRG results and the solid 
lines are visual guides. The mean-field results from reference ~\cite{Nabi-ArX16} are shown in the first two diagrams. Mott insulator (MI) lobes 
surrounded by a superfluid phase were found for any values of interaction parameters.}
\end{minipage}\hspace{2pc}%
\end{figure}
In order to characterize the magnetic order in spin-1 systems, it is important to calculate the $z$-component of the nematic order parameter 
$Q_{zz}=<\hat{S}_{z}^2>-\frac13<\mathbf{\hat{S}}^2>$, because in our case $Q_{xx}=Q_{yy}$. Then we obtain on-site spin isotropy if 
$Q_{zz}=0$ and spin anisotropy if $Q_{zz}\neq 0$, which is characteristic of the nematic order~\cite{Demler-PRL02,Forges-PRB13}. In Fig.~\ref{fig2},
we see that the $z$-component of the nematic order parameter vanishes for global densities $\rho\leq2$; hence for the hopping parameter $t/U_0=0.10$, 
the Mott insulator and superfluid regions will not be nematic. A longitudinal polar superfluid is characterized by the predominant $\sigma=0$ spin 
population ($N_0>N_+$) and this is spin isotropic $Q_{zz}=0$. The ground state remains a longitudinal polar superfluid for densities slightly greater
 than $\rho=2$, and then the $z$-component of the nematic order parameter jumps to a finite value, which coincides with the change in the spin 
population fractions, now  $Q_{zz}\neq 0$ and a non-nematic to nematic quantum phase transition has occurred. A transverse polar superfluid is 
characterized by the predominant $\sigma=\pm$ spin population ($N_\pm>N_0$), and it is spin anisotropic $Q_{zz}\neq 0$. Our numerical results 
indicate that the quantum phase transition from a longitudinal polar superfluid to a transverse polar superfluid is discontinuous. 
Note that a discontinuous phase transition between two superfluid phases for $^7$Li atoms in a three-dimensional optical lattice was discussed 
recently, but in this case the transition is driven by a quadratic Zeeman term, which competes with the ferromagnetic interaction between the 
spin-1 bosons~\cite{So-ArX16}.\par
\begin{figure}[t] 
\includegraphics[width=18pc]{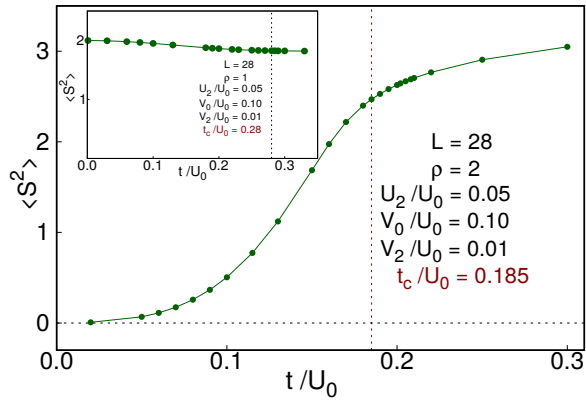} 
\caption{\label{fig5} The square of the local moment($<S^2>$) as a function of the hopping parameter $t/U_0$ for a global density 
$\rho = 2$. Here, we consider a chain with $L=28$ sites, and the interaction parameters are $U_2/U_0=0.05$, $V_0/U_0=0.1$, and $V_2/U_0=0.01$.
Inset: The same as the main plot but for $\rho = 1$. The vertical dashed line indicates the position of the Mott insulator-superfluid transition, which 
corresponds to $t_c/U_0=0.28$ and $t_c/U_0=0.185$ for the first and the second Mott lobe, respectively. The lines are visual guides.} 
\end{figure} 
In Fig. \ref{fig3}, we show the dependence of the system size on the chemical potential of a chain of spinor bosons with global density $\rho=1$ and 
interaction parameters $U_2/U_0 =0.05$, $V_0/U_0 =0.1$, and $V_2/U_0 =0.01$. The upper set of data shows the energy needed to add a particle to the system, 
while the lower one the energy needed to remove a particle, which were obtained with DMRG. The values at the thermodynamic limit ($N,L \rightarrow \infty $) 
were calculated by extrapolation, considering quadratic or linear dependence. The left panel shows the results for $t/U_0=0.10$ (Fig. \ref{fig3}a), 
while the right one shows the results for $t/U_0=0.28$ (Fig. \ref{fig3}b), and it is evident that the ground state changes as the hopping parameter increases. 
We obtain a gapped state for small values of the hopping with an integer number of particles per site; hence the ground state is a Mott insulator one, 
but as the hopping grows, we expect that spinor bosons will delocalize throughout the lattice and the ground state will be gapless (superfluid). This figure 
shows us that a quantum phase transition from a superfluid state to a Mott insulator one will take place as the hopping parameter decreases as the 
two- and three-body interactions between spinor bosons are considered. The above conclusion is expected when we are studying interacting bosons; 
however, the most interesting problem is related to the phase diagram, for which recent mean-field studies suggest that the even-odd asymmetry always 
happens~\cite{Nabi-ArX16}.\par
\begin{figure}[t]
\includegraphics[width=18pc]{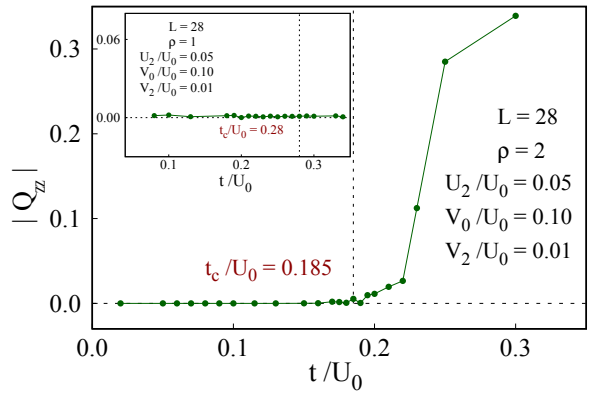}
\caption{\label{fig6} Evolution of the nematic structure factor $Q_{zz}$ as a function of the hopping parameter for $U_2/U_0=0.05$, $V_0/U_0=0.1$ and 
$V_2/U_0=0.01$, with global density $\rho=2$. Inset: The same as the main plot but for $\rho = 1$. The vertical dashed lines indicate the position of SF-MI 
transition, which corresponds to $t_c/U_0=0.28$ and $t_c/U_0=0.185$ for the first and the second Mott lobe, respectively. The lines are visual guides. }
\end{figure}
Although physical arguments and mean-field calculations lead to the conclusion that the ground state of interacting spinless or spinor bosons may be either 
Mott insulator or superfluid, with the superfluid surrounding the Mott insulator lobes, an accurate determination of the boundaries between these phases is 
an interesting problem that has been considered in many papers by means of different approaches and techniques. For instance, an accurate phase diagram for 
spinor bosons in one dimension with two-body (three-body) interactions between them was found using numerical approaches 
(DMRG and/or QMC)~\cite{Rizzi-PRL05,Batrouni-PRL09} (~\cite{Hincapie-PRA16}). We used the DMRG algorithm to find the phase diagram of a chain of 
spinor bosons under two- and three-body interactions by calculating the chemical potential to add or remove one particle at the thermodynamic limit 
for different values of the hopping, when the parameters $U_2$, $V_0<<U_0$, and $V_2=2\frac{U_2}{U_0}V_0$ are fixed. Mean-field phase diagrams for this 
problem were recently reported by Nabi and Basu~\cite{Nabi-ArX16}, and we show them in Fig. \ref{fig4}a and Fig. \ref{fig4}b (Rounded green and 
yellow regions, respectively). Note that the mean-field results suggest that the even-odd asymmetry of spinor bosons with two-body interactions is 
maintained with three-body interactions. We found that the extrapolated values from the DMRG results at the atomic limit are in agreement with 
the mean-field calculations and with our previous analysis; therefore, borders of the Mott lobes depend on the interaction parameters $U_2$ and $V_0$. In 
Fig. \ref{fig4}a, the interaction parameters are $U_2/U_0=0.05$, $V_0/U_0=0.1$, and $V_2/U_0=0.01$, and we obtained that for very small values of 
$ t/U_0 $, the mean-field and the results from DMRG data are close, but as the value of $ t / U_0 $ increases, the mean-field lobes close quickly, 
whereas the numerical ones close very slowly, moving the critical point to larger values of the hopping, and giving to the lobes an elongated 
shape regardless the global density. Note that the odd lobes are larger than the even lobe, indicating that the even-odd asymmetry suggested by the 
mean-field calculation is an artiifact of the method,  and that the three-body interactions contribute more to the localization of the particles 
for global densities greater than $\rho=1$. The first lobe remains almost the same, while the others grow, especially the Mott lobe with $\rho=3$, whose 
critical point is higher than the one for $\rho=2$. Maintaining the same spin-dependent two-body interaction value $U_2/U_0=0.05$ but increasing the 
spin-independent three-body interaction value $V_0/U_0=0.5$, we obtain the phase diagram shown in Fig. \ref{fig4}b. There, the borders of the first 
Mott lobe do not change at the atomic limit, and due to the small fluctuations of charge, the three-body interactions are of little importance, leading to a
first Mott lobe similar to the one in the previous figure, but with a slightly lower critical value. For the second lobe, we see that at the atomic 
limit the gap depends on the parameter $V_0$ and the three-body interactions become more important due to the increase in the global density, 
leading to the growth of second Mott lobe and to its critical point's moving to larger values of the hopping. Something similar happens for the third Mott 
lobe, this being the largest lobe for these parameters, with a critical point at $t_c (\rho=3)\approx0.32U_0$. Hence we conclude that for a fixed 
two-body spin interaction, the three-body interactions increase the Mott lobes for densities greater than $\rho=1$, and their critical points move 
to larger values of the hopping.\par 
Up to now, we have considered the spin two-body interaction equal to $U_2/U_0=0.05$ due to the phase diagrams reported by Nabi and Basu. However, the 
above value is small. Taking into account this fact, we consider a larger value $U_2/U_0=0.30$ and draw the phase diagram shown in Fig. \ref{fig4}c 
for $V_0/U_0=0.1$ and $V_2/U_0=0.06$. This time the odd Mott lobes are very small and the even one predominates in the phase diagram, showing the 
well-known even-odd asymmetry characteristic of spinor chains with only two-body interactions; i.e. for larger spin two-body interactions and 
smaller three-body interactions, the even-odd asymmetry is preserved. Comparing Fig. \ref{fig4}c with the results reported by 
Rizzi {\it et al.}~\cite{Rizzi-PRL05} for only two-body interactions, we observe that the first lobe is almost the same and the second one has a 
larger gap at the atomic limit, but the critical point is similar. At the atomic limit, the gap of the Mott lobe for $\rho=3$ is reduced by the 
interactions. The lower border rises due to $V_0/U_0$, while the upper border lowers due to $U_2/U_0$, but the smaller three-body interactions 
become important, elongating the lobe until it reaches a higher critical point $t_c (\rho=3)\approx0.24U_0$, which is higher than the case with only 
two-body interactions. Again, if we set the spin two-body interaction and increase the three-body interactions, the Mott lobes for $\rho>1$ grow and 
their critical points go to larger values (not shown).\par
Using different approaches and/or numerical techniques, several authors have shown that the magnetic properties of superfluid and Mott insulator 
phases of spin-1 boson systems are diverse and interesting. For instance, for an antiferromagnetic chain with two-body interactions, it has been 
suggested that the even Mott lobes exhibit a singlet and a nematic phase, whereas the odd lobes exhibit a dimerized 
phase~\cite{Rizzi-PRL05,Batrouni-PRL09}. When only three-body interactions are considered, it has been shown that the ground state is not composed of 
singlets at the second Mott lobe~\cite{Hincapie-PRA16}. The formation or lack thereof of singlets at the Mott lobes can be shown by calculating the square 
of the local moment($<S^2>$), as shown in Fig.  \ref{fig5} for a chain with $L=28$ sites, $U_2/U_0=0.05$, $V_0/U_0=0.1$, and $V_2/U_0=0.01$. For one spinor 
boson per site ($\rho=1$), we note that $<S^2>$ is finite and nonzero in both the Mott and the superfluid regions (see inset of Fig. \ref{fig5}). Obviously, no 
local singlets are formed, and the well-known dimerized pattern for this filling is obtained (not shown). Increasing the global density up to $\rho=2$, 
we observe that $<S^2>$ is non-zero for larger values of the hopping in the superfluid region. However, as the hopping decreases,  $<S^2>$ decreases 
monotonously, both in the superfluid region and in the Mott insulator lobe, reaching zero value at the atomic limit, where we have a singlet at each 
site. Note that the transition point shown in Fig. \ref{fig5} corresponds to a naive estimation based on the vanishing of the charge gap, which is not 
valid when a first-order transition takes place, as in the superfluid-Mott insulator transition for $\rho=2$.\par
\begin{figure}[t]
\includegraphics[width=18pc]{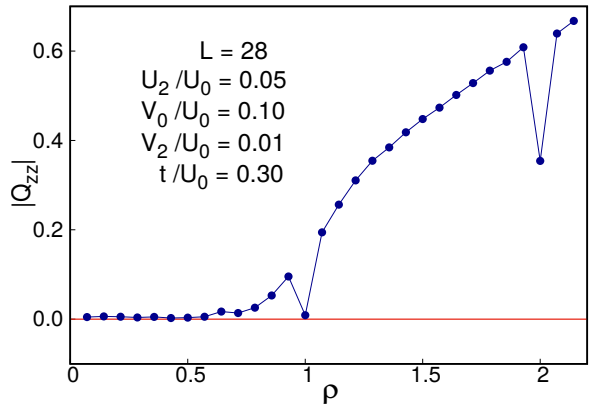}
\caption{\label{fig7} The nematic structure factor $Q_{zz}$ as a function of the global density $\rho$ at the superfluid region. Here the parameters are 
$t/U_0=0.30$, $U_2/U_0=0.05$, $V_0/U_0=0.1$ and $V_2/U_0=0.01$. The lines are visual guides.}
\end{figure}
For a fixed global density $\rho=2$ (main plot) and $\rho=1$ (inset), the evolution of the $z$-component of the nematic order parameter $Q_{zz}$ versus the 
hopping parameter is shown in Fig. \ref{fig6} from deep in the Mott lobes to the superfluid phase. We found that in the Mott insulator lobes, the spin 
remains isotropic: $Q_{zz}=0$. This result is expected for $\rho=1$, because the ground state is dimerized; however, for $\rho=2$, our finding is curious, 
because the possibility of a singlet-nematic transition in even lobes has been suggested by several authors~\cite{Batrouni-PRL09,Forges-PRB13}. Our ground 
state is composed of singlets, and the $<S^2>$ non-zero values are due to the quantum fluctuations. The inset of Fig. \ref{fig6} tells us that the ground 
state is not nematic for either the Mott insulator or the superfluid regions when one boson per site is considered, but for a global density $\rho=2$, the 
ground state is not nematic in the Mott lobe, but shortly after the transition the spin anisotropy ($Q_{zz}\neq 0$) grows continuously starting at zero 
in the superfluid region, this being a behavior similar to that reported in 2D for larger values of the two-body spin dependent 
interaction~\cite{Forges-PRB13}. In the superfluid region and at $t/U_0=0.30$, we see that the ground state is non-nematic and nematic for densities 
$\rho=1$ and $\rho=2$, respectively. Due to this fact, we calculated $Q_{zz}$ as a function of the global density in the superfluid region, and the results 
are shown in Fig. \ref{fig7}. We obtained that the ground state is a longitudinal polar superfluid for small densities, but for densities larger than 
$\rho\approx 0.55$, a continuous quantum phase transition takes place and the ground state becomes a transverse polar superfluid except for $\rho=1$, for 
which the ground state is always a longitudinal polar superfluid. Note that the evolution of the spin anisotropy is monotonous except for the integer 
densities, for which it falls. Recently, direct experimental evidence of spin-nematic ordering in a spin-1 Bose-Einstein condensate of sodium 
atoms with antiferromagnetic interactions has been shown. We believe that based on these measurements it could be possible to identify multi-body interaction 
effects~\cite{Zibold-PRA16}.

\section{Conclusions\label{conclu}}
To summarize, in the present paper, we used the density matrix renormalization group to study the ground state of spin-1 bosons loaded into a 
one-dimensional optical lattice interacting via two- and three-body spin dependent and independent interaction terms, considering only 
antiferromagnetic interaction. The ground state can be Mott insulator or superfluid, and the border between them was found for different sets of 
the interaction parameters, showing that the well-known even-odd asymmetry for the case with only two-body interactions does not obtain and depends on 
the relative values of the two- and three-body interactions, contradicting the mean-field results of Nabi and Basu, who suggested that the 
even-odd asymmetry persists~\cite{Nabi-ArX16}. We found that the three-body interactions do not alter the kind of transition, i.e. the first (second) 
Mott insulator-superfluid transition is second- (first-) order. Regardless of the strength of the two-body spin-dependent interactions, between the second 
and the third Mott lobe a quantum phase transition from a longitudinal polar superfluid to a transverse polar superfluid takes place. Exploring the 
magnetic properties of the first two Mott lobes, we found that they are spin isotropic, the second Mott lobe being composed of singlets, preventing the 
possibility of a nematic phase inside, which has been suggested for the case with only two-body interactions in one and two dimensions. For one boson 
per site the superfluid is always longitudinal. Our numerical results suggest that the longitudinal-transverse polar superfluid transition can be either
continuous or discontinuous, and the critical density for which it happens decreases as the hopping parameter increases, in a way similar to quantum transitions
between superfluid states in a spinor system with a quadratic Zeeman potential~\cite{So-ArX16}.

\section*{Acknowledgments}
The authors are thankful for the support of Departamento Administrativo de Ciencia, Tecnolog\'{\i}a e Innovaci\'on (COLCIENCIAS) 
(Grant No. FP44842-057-2015). J.S.-V. is grateful for the hospitality of the ICTP, where part of this work was done.

\section*{References}


\end{document}